\documentclass[aps,prd,showpacs,preprint]{revtex4}

\usepackage{amsmath}
\usepackage{amssymb}

\begin{document}

\title{Minimum mass-radius ratio for charged gravitational objects}

\author{C. G. B\"ohmer}
\email{christian.boehmer@port.ac.uk}
\affiliation{Institute of Cosmology \& Gravitation,
             University of Portsmouth, Portsmouth PO1 2EG, UK}
\affiliation{ASGBG/CIU, Department of Mathematics, Apartado Postal C-600, 
             University of Zacatecas (UAZ), Zacatecas, Zac 98060, Mexico}

\author{T. Harko}
\email{harko@hkucc.hku.hk} 
\affiliation{Department of Physics and Center for Theoretical
             and Computational Physics, The University of Hong Kong,
             Pok Fu Lam Road, Hong Kong}

\date{March 6, 2007}

\begin{abstract}
We rigorously prove that for compact charged general relativistic
objects there is a lower bound for the mass-radius ratio. This
result follows from the same Buchdahl type inequality for charged
objects, which has been extensively used for the proof of the existence 
of an upper bound for the mass-radius ratio. The effect of the vacuum
energy (a cosmological constant) on the minimum mass is also taken into 
account. Several bounds on the total charge, mass and the vacuum energy 
for compact charged objects are obtained from the study of the Ricci 
scalar invariants. The total energy (including the gravitational one) 
and the stability of the objects with minimum mass-radius ratio is also 
considered, leading to a representation of the mass and radius of the 
charged objects with minimum mass-radius ratio in terms of the charge 
and vacuum energy only.
\end{abstract}

\maketitle

\section{Introduction}

The bag models of hadrons~\cite{bag}, proposed in the 1970's, have had a
remarkable phenomenological success (see~\cite{HoTo96} and~\cite{Bu05} for
reviews and recent developments). In these models, hadrons consist of free
(or only weekly interacting) quarks, which are confined to a finite region
of space, called the bag. The confinement is not a dynamical one, but it is
put in by hand, imposing some appropriate boundary conditions. The bag is
stabilised by a term of the form $g_{\mu \nu }B$, which is added to the
energy momentum tensor $T_{\mu \nu }$ inside the bag, which thus takes the
form $T_{\mu \nu }=T_{\mu \nu }^{(\mathrm{fields})}+g_{\mu \nu }B$. By
recalling the energy-momentum tensor of a perfect fluid in its rest-frame, $%
T^{\mu}_{\nu }=\mathrm{diag}\left(\epsilon,-p,-p,-p\right)$, where $\epsilon $
is the energy density and $p$ is the thermodynamic pressure, it immediately
follows that the bag constant $B$ is immediately interpreted as positive
contribution to the energy density $\epsilon $ and a negative contribution
to the pressure $p$ inside the bag. Equivalently, we may attribute a term $%
-g_{\mu \nu }B$ to the region outside the bag. This leads to a picture of a
non-trivial vacuum with a negative energy density $\epsilon _{vac}=-B$ and a
positive pressure $p_{vac}=+B$. The stability of the hadron then results
from balancing this positive vacuum pressure with the pressure caused by the
quarks inside the bag~\cite{Bu05}.

Therefore, quark bag models in the theories of strong interactions assume
that the breaking of physical vacuum takes place inside hadrons. As a result
the vacuum energy densities inside and outside a hadron become essentially
different and the vacuum pressure $B$ on a bag wall equilibrates the
pressure of quarks thus stabilising the system. The MIT bag model says
nothing about the origin of the non-trivial vacuum, but treats $B$ as a free
parameter. Assuming a static spherical bag of radius $R$, the mass of the
hadron is given by the sum $E_{BM}=4\pi BR^{3}/3-z_{0}/R+\sum_{q}x_{q}/R+...$, 
where the first term corresponds to the volume energy, required to replace
the non-trivial vacuum by the trivial one inside the bag, the second term
parameterise the finite part of the zero-point energy of the bag and the
third term is the sum of the rest and kinetic energy of the quarks~\cite{Bu05}.

The finite electron self-energy is a puzzling problem in both
quantum theory and classical theory. Quantum electrodynamics, with
its remarkable predictive power, fails to explain the origin of
the finite electron mass, and none of the proposed regularisation
schemes have succeeded in predicting the observed mass. On the
other hand, a point charge is incompatible with classical
electrodynamics, because it has the self-energy and stability
problems. An electron of finite radius was proposed by Abraham and
Lorentz, with the particle radius equal to $R=Q^{2}/M$, where $Q$
and $M$ are the charge and the mass of the particle, respectively.
This relation has been obtained by assuming that the
electromagnetic potential energy of the particle $Q^{2}/R$ is
equal to its mass $M$, according to the mass-energy equivalence
law. However, an extended charge distribution interacting with
itself cannot be stable and non-electromagnetic forces are needed
to prevent the electron from exploding. Such cohesive
non-electromagnetic forces were suggested by Poincar\'{e}, and are
called Poincar\'{e} stresses~\cite{Po03}.

On the other hand, the Einstein-Maxwell field equations of general
relativity can be used to construct a Lorentz model of an electron
as an extended body consisting of pure charge and no matter and
electromagnetic mass models for static spherically symmetric
charged fluid distributions have been extensively studied
\cite{elec}. The Poincar\'{e} stresses are explained as due to
vacuum polarisation, the vacuum energy density $\rho _{V} $ and
the vacuum pressure $p_{V}$ satisfying an equation of state of
the form $\rho _{V}+p_{V}=0$, where in general the vacuum energy density $%
\rho _{V}>0$ and the pressure $p_{V}<0$. This type of equation of state
implies that the matter distribution under consideration is in tension, in a
state known as ``false vacuum'' or ``degenerate vacuum''. The gravitational
blue-shift of light is explained as due to repulsive gravitation produced by
the negative gravitational mass of the polarised vacuum. In the context of
general relativity, the electron, modelled as a spherically symmetric
charged distribution of matter, must contain some negative rest mass if its
radius is not larger than $10^{-16}$ cm. In some extended electron models,
the negative energy density distributions result from the requirement that
the total mass of these models remains constant in the limit of a point
particle.

The mass-radius-charge relation for elementary particles,compact
astrophysical objects or black holes plays an important role in
many physical processes.  The pressure and the density of the
matter inside the stars are large, and the gravitational field is
intense. This indicates that electric charge and a strong electric
field may also be present. The effect of electric charge in
compact stars assuming that the charge distribution is
proportional to the mass density was studied in~\cite{el1}. In
order to see any appreciable effect on the phenomenology of the
compact stars, the electric fields have to be huge ($10^{21}$
V/m), which implies that the total charge is $Q\approx 10^{20}$
Coulomb. The star can then collapse to form a charged black hole.
Charged stars have the potential of becoming charged black holes
or even naked singularities. A set of numerical solutions of the
Tolman-Oppenheimer-Volkov equations that represents spherical
charged compact stars in hydrostatic equilibrium were obtained in~\cite{el2}. 
Charged boson stars in scalar-tensor gravitational
theories have been studied in~\cite{WhTo99}. In these models there
is a maximum charge to mass ratio for the bosons above which the
weak field solutions are not stable. This charge limit can be
greater than the general relativistic limit for a wide class of
scalar-tensor theories. The black hole formation in the head-on
collision of ultra-relativistic charges was studied in~\cite{YoMa06}. 
The formation of the apparent horizon was analysed and
a condition was obtained, indicating that a critical value of the
electric charge is necessary for black hole formation to take
place. By evaluating this condition for characteristic values at
the LHC, it was found that the presence of the charge decreases
the black hole production rate in accelerators. Mass-charge limits
are important for the study of the quasi-local energy measured by
observers who are moving around in the space-time. The quasi-local
formalism for gravitational energy was extended in~\cite{BoMa99}
to include electromagnetic and dilaton fields and to also allow
for spatial boundaries that are not orthogonal to the foliation of
the space-time. The distribution of energy around
Reissner-Nordstr\"om and naked black holes was investigated as
measured by both static and infalling observers. The study of
naked black holes reveals an alternate characterisation of this
class of space-times in terms of the quasi-local energies.

The observations of high redshift supernovae~\cite{Pe99} and the
Boomerang/Maxima data~\cite{Ber00}, showing that the location of the first
acoustic peak in the power spectrum of the microwave background radiation is
consistent with the inflationary prediction $\Omega =1$, have provided
compelling evidence for a net equation of state of the cosmic fluid lying in
the range $-1\leq w=p/\rho <-1/3$. To explain these observations, two dark
components are invoked: the pressure-less cold dark matter (CDM) and the
dark energy (DE) with negative pressure. CDM contributes $\Omega_{m}\sim 0.25$, 
and is mainly motivated by the theoretical interpretation of the
galactic rotation curves and large scale structure formation. DE provides $%
\Omega _{DE}\sim 0.7$ and is responsible for the acceleration of the distant
type Ia supernovae. The best candidate for the dark energy is the
cosmological constant $\Lambda $, which is usually interpreted physically as
a vacuum energy. Its size is of the order $\Lambda \approx 3\times 10^{-56}$
cm$^{-2}$~\cite{PeRa03}.

However, the WMAP data also allow the possibility that the
Universe may be slightly above/below the $\Lambda $CDM model, in
the so called phantom region (see~\cite{od1} and references
therein). In the phantom scenario the acceleration of the Universe
is explained by the presence of some phantom matter, with negative
energy density. The similarity of phantom matter with quantum CFT
indicates that the phantom scalar may be the effective description
for some quantum field theory~\cite{od1}. For phantom/tachyonic
matter the standard energy conditions of general relativity, the
null energy condition (NEC) $\rho +p\geq 0$, the weak energy
condition (WEC) $\rho \geq 0 $ and $\rho +p\geq 0$, the strong
energy condition (SEC) $\rho +3p\geq 0$
and $\rho +p\geq 0$ and dominant energy condition (DEC) $\rho \geq 0$ and $%
\rho \pm p\geq 0$ are violated~\cite{kahya,od2}. Such a model naturally admits two
de Sitter phases where the early universe inflation is produced by quantum
effects and the late time accelerating universe is caused by
phantom/tachyon.  The typical final state of a dark energy universe where a
dominant energy condition is violated is a finite-time, sudden future
singularity (a big rip). For a number of dark energy universes (including
scalar phantom and effective phantom theories as well as specific
quintessence models) the quantum effects play the dominant role near a big
rip, driving the universe out of a future singularity~\cite{od3}. Black hole
mass loss due to phantom accretion is very different from the standard
general relativistic case: masses do not vanish to zero due to the transient
character of the phantom evolution stage~\cite{od3}.

By using the static spherically symmetric gravitational field equations
Buchdahl~\cite{Bu59} has obtained an absolute constraint of the maximally
allowable mass $M$ and radius $R$ for isotropic fluid spheres of the form $%
2M/R\leq 8/9$ (where natural units $c=G=1$ have been used).

The existence of the cosmological constant modifies the allowed ranges for
various physical parameters, like, for example, the maximum mass of compact
stellar objects, thus leading to a modification of the ``classical'' Buchdahl
limit~\cite{MaDoHa00}, for the effect of anisotropy, see e.g.~\cite{BoHa06}.

The maximum allowable mass-radius ratio in the case of stable charged
compact general relativistic objects was obtained in~\cite{MaDoHa01}, by
generalising to the charged case the methods used for neutral stars by
Buchdahl~\cite{Bu59} and Straumann~\cite{St84}.

On the other hand, we cannot exclude \textit{a priori} the possibility that
the cosmological constant, as a manifestation of vacuum energy, may play an
important role not only at galactic or cosmological scales, but also at the
level of elementary particles (the very successful phenomenological bag
model of hadrons requires the existence of the vacuum energy inside and
outside strongly interacting particles). With the use of the generalised
Buchdahl identity~\cite{MaDoHa00}, it can be rigorously proven that the
existence of a non-negative $\Lambda $ imposes a lower bound on the mass $M$
and density $\rho $ of general relativistic objects of radius $R$, which is
given by~\cite{BoHa05}
\begin{equation}
2M\geq \frac{8\pi \Lambda }{6}R^{3},\qquad \rho =\frac{3M}{4\pi R^{3}}\geq
\frac{\Lambda }{2}=:\rho _{\min }.  \label{minm}
\end{equation}

Therefore, the existence of the cosmological constant implies the existence
of an absolute minimum mass and density in the universe. No object present
in relativity can have a density that is smaller than $\rho _{\min }$. For $%
\Lambda >0$ this result also implies a minimum density for stable
fluctuations in energy density.

It is the purpose of the present paper to consider the problem of the
existence of a minimum mass-radius ratio for compact electrically charged
general relativistic objects. We rigorously prove that a lower bound for the
ratio $M/R$ does exist for charged objects with non-zero electric charge $Q$. 
This result follows from the same Buchdahl type inequality which has been
extensively used for the proof of the existence of an upper bound for the
mass-radius ratio.

The present paper is organised as follows. The generalised Buchdahl
inequality for charged objects in the presence of a vacuum energy (a
cosmological constant) is derived in Section II. In Section III we obtain
some bounds on the total charge and mass of compact charged objects from the
study of the Ricci scalar invariants. The total energy (including the
gravitational one) and the stability of the objects with minimum mass-radius
ratio is considered in Section IV. We discuss and conclude our results in
Section V.

Throughout this paper we use the Landau-Lifshitz conventions~\cite{LaLi76}
for the metric signature $\left( +,-,-,-\right) $ and for the field
equations, and a system of units with $c=G=\hbar =1$.

\section{Generalised Buchdahl inequality for charged objects}

For a static general relativistic spherically symmetric configuration the
interior line element is given by
\begin{equation}
ds^{2}=e^{\nu \left( r\right) }dt^{2}-e^{\lambda \left( r\right)}dr^{2}
-r^{2}\left( d\theta^{2}+\sin^{2}\negmedspace\theta\, d\varphi ^{2}\right) .
\end{equation}

The properties of a charged compact general relativistic object can be
completely described by the structure equations, which are given by
\begin{align}
      \frac{dm}{dr}&=4\pi \rho r^{2}+\frac{Q}{r}\frac{dQ}{dr},  
      \label{1} \\[1ex]
      \frac{dp}{dr}&=-\frac{\left( \rho +p\right) \left[ m+4\pi r^{3}
      \left( p-\frac{2B}{3}\right) -\frac{Q^{2}}{r}\right] }
      {r^{2}\left( 1-\frac{2m}{r}+\frac{Q^{2}}{r^{2}}-\frac{8\pi }{3}Br^{2}\right)}
      +\frac{Q}{4\pi r^{4}}\frac{dQ}{dr},  
      \label{2} \\[1ex]
      \frac{d\nu }{dr}&=\frac{2\left[ m+4\pi r^{3}\left( p-\frac{2B}{3}\right)-
      \frac{Q^{2}}{r}\right] }{r^{2}\left( 1-\frac{2m}{r}+\frac{Q^{2}}{r^{2}}-
      \frac{8\pi }{3}Br^{2}\right) },  
      \label{3}
\end{align}
where $\rho \left( r\right) $ is the energy density of the matter, $p\left(
r\right) $ is the thermodynamic pressure, $m(r)$ is the mass and
\begin{equation}
Q(r)=4\pi \int_{0}^{r}e^{\frac{\nu +\lambda }{2}}r'^{2}j^{0}dr',
\end{equation}
is the electric charge inside radius $r$, respectively. The electric current
inside the charged object is given by $j^{\mu }=\left( j^{0},0,0,0\right) $.
By analogy with the bag model of hadrons we also assume the presence of an
effective constant vacuum energy density $B$ (a cosmological constant)
inside and outside the charged object. Eqs.~(\ref{1})--(\ref{3}) represent
the generalisation of the structure equations for general relativistic
static charged objects, introduced for the first time in~\cite{Be71}, by
taking into account the existence of a non-zero vacuum energy.

Generally $p$ and $\rho $ are related by an equation of state of the form
$\rho=\rho(p)$. The structure equations Eqs.~(\ref{1})--(\ref{3})
must be considered together with the boundary conditions $p(R)=0$, 
$p(0)=p_{c}$, $\rho_{c}=\rho(p=0)$ and $Q(0)=0$, where $\rho_{c}$, $p_{c}$
are the central density and pressure, respectively.

With the use of Eqs.~(\ref{1})--(\ref{3}) it is easy to show that the
function $\zeta =\exp \left( \nu /2\right) >0$, $\forall r\in \lbrack 0,R]$,
obeys the equation
\begin{equation}
\sqrt{1-\frac{2m}{r}+\frac{Q^{2}}{r^{2}}-\frac{8\pi }{3}Br^{2}}\frac{1}{r}%
\frac{d}{dr}\left[ \sqrt{1-\frac{2m}{r}+\frac{Q^{2}}{r^{2}}-\frac{8\pi }{3}%
Br^{2}}\frac{1}{r}\frac{d\zeta }{dr}\right] =\frac{\zeta }{r}\left[ \frac{d}{%
dr}\frac{m}{r^{3}}+\frac{Q^{2}}{r^{5}}\right] .  \label{4}
\end{equation}

For $Q=0$ and $B=0$ we obtain the equation considered in~\cite{St84}. Since
the density $\rho $ does not increase with increasing $r$, the mean density
of the matter $\langle\rho\rangle=3m(r) /4\pi r^{3}$ inside radius $r$ does
not increase either. Therefore we assume that inside a compact general
relativistic object the condition
\begin{equation}
\frac{d}{dr}\frac{m}{r^{3}}<0,
\end{equation}
holds, independently of the equation of state of dense matter and of the
electric charge distribution inside the object.

By defining a new function
\begin{equation}
\eta (r)=\int_{0}^{r}\frac{r^{\prime }}{\sqrt{1-\frac{2m\left( r^{\prime
}\right) }{r^{\prime }}+\frac{Q^{2}\left( r^{\prime }\right) }{r^{\prime 2}}}%
-\frac{8\pi }{3}Br^{\prime 2}}\left[ \int_{0}^{r^{\prime }}\frac{Q^{2}\left(
r^{\prime \prime }\right) \zeta \left( r^{\prime \prime }\right) }{r^{\prime
\prime 5}\sqrt{1-\frac{2m\left( r^{\prime \prime }\right) }{r^{\prime \prime
}}+\frac{Q^{2}\left( r^{\prime \prime }\right) }{r^{\prime \prime 2}}-\frac{%
8\pi }{3}Br^{\prime \prime 2}}}dr^{\prime \prime }\right] dr^{\prime },
\label{5}
\end{equation}
denoting $\Psi =\zeta -\eta $, and introducing a new independent variable $%
\xi \left( r\right) $ by means of the transformation~\cite{MaDoHa01,St84}
\begin{equation}
\xi \left( r\right) =\int_{0}^{r}r^{\prime }\left[ 1-\frac{2m(r^{\prime })}{%
r^{\prime }}+\frac{Q^{2}\left( r^{\prime }\right) }{r^{\prime 2}}-\frac{8\pi
}{3}Br^{\prime 2}\right] ^{-\frac{1}{2}}dr^{\prime },
\end{equation}
from Eq.~(\ref{5}) we obtain the basic result that inside all stable stellar
type charged general relativistic matter distributions the condition
\begin{equation}
\frac{d^{2}\Psi }{d\xi ^{2}}<0,
\end{equation}
must hold for all $r\in \left[ 0,R\right] $. Using the mean value theorem
\cite{St84} we conclude that
\begin{equation}
\frac{d\Psi }{d\xi }\leq \frac{\Psi \left( \xi \right) -\Psi (0)}{\xi },
\end{equation}
or, taking into account that $\Psi (0)>0$ it follows that,
\begin{equation}
\Psi ^{-1}\frac{d\Psi }{d\xi }\leq \frac{1}{\xi }.  \label{6}
\end{equation}

In the following we denote
\begin{equation}
\alpha (r)=1-\frac{Q^{2}(r)}{2m(r)r}+\frac{4\pi }{3}B\frac{r^3}{m(r)}.
\end{equation}

In the initial variables the inequality (\ref{6}) takes the form
\begin{multline}  \label{7}
\frac{1}{r}\sqrt{ 1-\frac{2\alpha \left( r\right) m(r)}{r}}\left\{ \frac{1%
}{2}\frac{d\nu }{dr}e^{\frac{\nu (r)}{2}}-\frac{r}{\sqrt{1-\frac{2\alpha
\left( r\right) m(r)}{r}}}\int_{0}^{r}\frac{Q^{2}\left( r^{\prime }\right)
e^{\frac{\nu \left( r^{\prime }\right) }{2}}}{r^{\prime 5}\sqrt{1-\frac{%
2\alpha \left( r^{\prime }\right) m(r^{\prime })}{r^{\prime }}}}dr^{\prime
}\right\} \leq \\
\frac{e^{\frac{\nu (r)}{2}}-\int_{0}^{r}r^{\prime }\left[ 1-\frac{2\alpha
\left( r^{\prime }\right) m(r^{\prime })}{r^{\prime }}\right] ^{-\frac{1}{2}%
}\left\{ \int_{0}^{r^{\prime }}\left[ 1-\frac{2\alpha \left( r^{\prime
\prime }\right) m(r^{\prime \prime })}{r^{\prime \prime }}\right] ^{-\frac{1%
}{2}}\frac{Q^{2}\left( r^{\prime \prime }\right) e^{\frac{\nu \left(
r^{\prime \prime }\right) }{2}}}{r^{\prime \prime 5}}dr^{\prime \prime
}\right\} dr^{\prime }}{\int_{0}^{r}r^{\prime }\left[ 1-\frac{2\alpha \left(
r^{\prime }\right) m(r^{\prime })}{r^{\prime }}\right] ^{-\frac{1}{2}%
}dr^{\prime }}.
\end{multline}

For any stable charged compact objects $m/r^{3}$ does not increase outwards.
We suppose that for all $r^{\prime }\leq r$ we have
\begin{equation}
\frac{\alpha \left( r^{\prime }\right) m(r^{\prime })}{r^{\prime }}\geq
\frac{\alpha \left( r\right) m(r)}{r}\left( \frac{r^{\prime }}{r}\right)
^{2},
\end{equation}
or, equivalently,
\begin{equation}
\frac{2m\left( r^{\prime }\right) }{r^{\prime }}-\frac{2m(r)}{r}\left( \frac{%
r^{\prime }}{r}\right) ^{2}\geq \frac{Q^{2}(r^{\prime })}{r^{\prime 2}}-%
\frac{Q^{2}(r)}{r^{2}}\left( \frac{r^{\prime }}{r}\right) ^{2}.
\end{equation}

We also assume that inside the compact stellar object the charge $Q(r)$ satisfies
the general condition
\begin{equation}  \label{8}
\frac{Q^2(r^{\prime \prime })e^{\frac{\nu \left( r^{\prime \prime }\right) }{2}%
}}{r^{\prime \prime 5}}\geq \frac{Q^2(r^{\prime })e^{\frac{\nu \left(
r^{\prime }\right) }{2}}}{r^{\prime 5}}\geq \frac{Q^2(r)e^{\frac{\nu (r)}{2}}}{%
r^{5}},r^{\prime \prime }\leq r^{\prime }\leq r.
\end{equation}

Therefore, we can evaluate the terms in Eq.~(\ref{7}) as follows. For the
term in the denominator of the right hand side of Eq.~(\ref{7}) we obtain:
\begin{equation}
\left\{ \int_{0}^{r}r^{\prime }\left[ 1-\frac{2\alpha \left( r^{\prime
}\right) m\left( r^{\prime }\right) }{r^{\prime }}\right] ^{-\frac{1}{2}%
}dr^{\prime }\right\} ^{-1}\leq \frac{2\alpha (r)m(r)}{r^{3}}\left[ 1-\sqrt{%
1-\frac{2\alpha (r)m(r)}{r}}\right] ^{-1}.  \label{9}
\end{equation}
For the second term in the bracket of the left hand side of Eq.~(\ref{7}) we
have
\begin{multline}
\int_{0}^{r}\left[ 1-\frac{2\alpha \left( r^{\prime }\right) m\left(
r^{\prime }\right) }{r^{\prime }}\right] ^{-\frac{1}{2}}\frac{Q^{2}\left(
r^{\prime }\right) e^{\frac{\nu \left( r^{\prime }\right) }{2}}}{r^{\prime 5}%
}dr^{\prime }\\ \geq \label{10}
\frac{Q^{2}(r)e^{\frac{\nu (r)}{2}}}{r^{5}}\int_{0}^{r}\left[ 1-\frac{%
2\alpha \left( r\right) m(r)}{r}\left( \frac{r^{\prime }}{r}\right) ^{2}%
\right] ^{-\frac{1}{2}}dr^{\prime }\\ =
\frac{Q^{2}(r)e^{\frac{\nu (r)}{2}}}{r^{5}}\left[ \frac{2\alpha (r)m(r)}{%
r^{3}}\right] ^{-\frac{1}{2}}\arcsin \left[ \sqrt{\frac{2\alpha (r)m(r)}{r}}%
\right] .
\end{multline}

The second term in the nominator of the right hand side of Eq.~(\ref{7}) can
be evaluated as
\begin{multline}
\int_{0}^{r}r^{\prime }\left[ 1-\frac{2\alpha \left( r^{\prime }\right)
m\left( r^{\prime }\right) }{r^{\prime }}\right] ^{-\frac{1}{2}}\left\{
\int_{0}^{r^{\prime }}\left[ 1-\frac{2\alpha \left( r^{\prime \prime
}\right) m\left( r^{\prime \prime }\right) }{r^{\prime \prime }}\right] ^{-%
\frac{1}{2}}\frac{Q^{2}\left( r^{\prime \prime }\right) e^{\frac{\nu \left(
r^{\prime \prime }\right) }{2}}}{r^{\prime \prime 5}}dr^{\prime \prime
}\right\} dr^{\prime } \\ \geq \label{11}
\int_{0}^{r}r^{\prime }\left[ 1-\frac{2\alpha \left( r^{\prime }\right)
m\left( r^{\prime }\right) }{r^{\prime }}\right] ^{-\frac{1}{2}}\frac{%
Q^{2}\left( r^{\prime }\right) e^{\frac{\nu \left( r^{\prime }\right) }{2}}}{%
r^{\prime 4}}\left[ \frac{2\alpha (r^{\prime })m(r^{\prime })}{r^{\prime }}%
\right] ^{-\frac{1}{2}}\arcsin \left[ \sqrt{\frac{2\alpha (r^{\prime
})m(r^{\prime })}{r^{\prime }}}\right] dr^{\prime } \\ \geq
\frac{Q^{2}(r)e^{\frac{\nu (r)}{2}}}{r^{5}}\int_{0}^{r}r^{\prime 2}\left[
1-\frac{2\alpha (r)m(r)}{r^{3}}r^{\prime 2}\right] ^{-\frac{1}{2}}\left[
\frac{2\alpha (r)m(r)}{r^{3}}r^{\prime 2}\right] ^{-\frac{1}{2}}\arcsin %
\left[ \sqrt{\frac{2\alpha (r)m(r)}{r^{3}}}r^{\prime }\right] dr^{\prime }\\=
\frac{Q^{2}(r)e^{\frac{\nu (r)}{2}}}{r^{\frac{1}{2}} (2\alpha (r)m(r))^{\frac{3}{2}}}
\left\{ \sqrt{\frac{2\alpha (r)m(r)}{r}}-\sqrt{1-%
\frac{2\alpha (r)m(r)}{r}}\arcsin \left[ \sqrt{\frac{2\alpha (r)m(r)}{r}}%
\right] \right\} .
\end{multline}

In order to obtain the inequality (\ref{11}) we have also used the property
of monotonic increase in the interval $x\in \left[ 0,1\right] $ of the
function $\arcsin x/x$.

Using Eqs.~(\ref{9})--(\ref{11}), Eq.~(\ref{7}) becomes:
\begin{multline}
\left[ 1-\sqrt{1-\frac{2\alpha (r)m(r)}{r}}\right] 
\frac{m(r)+4\pi r^{3}\left( p-\frac{2}{3}B\right) -\frac{Q^{2}}{r}}{r^{3}
\sqrt{1-\frac{2\alpha (r)m(r)}{r}}} \\
\leq \frac{2\alpha (r)m(r)}{r^{3}}+\frac{Q^{2}}{r^{4}} 
\left\{ \frac{\arcsin \left[ \sqrt{\frac{2\alpha (r)m(r)}{r}}\right] }
{\sqrt{\frac{2\alpha (r)m(r)}{r}}}-1\right\}. \label{12}
\end{multline}

The Buchdahl type inequality given by Eq.~(\ref{12}) is valid for all $r$
inside the electrically charged object. It naturally leads to the existence
of a maximum mass-radius ratio for general relativistic objects.

Consider first the neutral case $Q=0$ and assume that the vacuum energy is
zero, $B=0$. We assume that at the surface of the compact object, defined by
a radius $r=R$, the thermodynamical pressure $p$ vanishes, $p(R)=0$. By
evaluating (\ref{12}) for $r=R$ we obtain $\left( 1-2M/R\right) ^{-1/2}\leq 2%
\left[ 1-\left( 1-2M/R\right) ^{-1/2}\right] ^{-1},$ leading to the
well-known result $2M/R\leq 8/9$~\cite{Bu59,St84}. The maximum mass-radius
ratio for charged object, representing the generalisation to the charged
case of the Buchdahl limit, was considered, and extensively discussed, in
the case of a vanishing vacuum energy $B=0$, in~\cite{MaDoHa01}.

\section{Minimum mass-radius ratio for charged general relativistic objects}

Eq.~(\ref{12}) also implies the existence of a minimum mass-radius ratio
for compact charged general relativistic objects. This can be shown as
follows. For small values of the argument the function $\arcsin x/x-1$ can
be approximated as $\arcsin x/x-1\approx x^{2}/6$. Therefore, at the vacuum
boundary $r=R$ of the charged object, Eq.~(\ref{12}) can be written in an
equivalent form as
\begin{equation}
\sqrt{1-\frac{2M}{R}+\frac{Q^{2}}{R^{2}}-\frac{8\pi }{3}BR^{2}}\geq \frac{M-%
\frac{Q^{2}}{R}-\frac{8\pi }{3}BR^{3}}{3M-2\frac{Q^{2}}{R}+\frac{Q^{2}}{%
6R^{2}}\left( 2M-\frac{Q^{2}}{R}+\frac{8\pi }{3}BR^{3}\right) }.  \label{an2}
\end{equation}

By introducing a new variable $u$ defined as
\begin{equation}
u=\frac{M}{R}-\frac{Q^{2}}{2R^{2}}+\frac{4\pi }{3}BR^{2},
\end{equation}
Eq.~(\ref{an2}) takes the form
\begin{equation}
\sqrt{1-2u}\geq \frac{u-a}{bu-a},  \label{an3}
\end{equation}
where we denoted$\bigskip $%
\begin{equation}
a=\frac{Q^{2}}{2R^{2}}+4\pi BR^{2},
\end{equation}
and \bigskip
\begin{equation}
b=3+\frac{Q^{2}}{3R^{2}},
\end{equation}
respectively. Then, by squaring, we can reformulate the condition given by
Eq.~(\ref{an3}) as
\begin{equation}
u\left[ 2b^{2}u^{2}-\left( b^{2}+4ab-1\right) u+2a\left( a+b-1\right) \right]
\leq 0,
\end{equation}
or, equivalently,
\begin{equation}
u\left( u-u_{1}\right) \left( u-u_{2}\right) \leq 0,  \label{cond2}
\end{equation}
where
\begin{equation}
u_{1}=\frac{b^{2}+4ab-1-\left( 1-b\right) \sqrt{(1+b)^{2}-8ab}}{4b^{2}},
\end{equation}
and
\begin{equation}
u_{2}=\frac{b^{2}+4ab-1+\left( 1-b\right) \sqrt{(1+b)^{2}-8ab}}{4b^{2}},
\end{equation}
respectively.

Since $u\geq 0$, Eq.~(\ref{cond2}) is satisfied if $u\leq u_{1}$ and $u\geq
u_{2}$, or $u\geq u_{1}$ and $u\leq u_{2}$. However, the condition $u\geq
u_{1}$ contradicts the upper bound which follows from Eq.~(\ref{12}), and
which has been discussed in detail in~\cite{MaDoHa01}. Therefore, Eq.~(\ref{cond2}) 
is satisfied if and only if for all values of the physical
parameters the condition $u\geq u_{2}$ holds. This is equivalent to the
existence of a minimum bound for the mass-radius ratio of compact
anisotropic objects, which is given by
\begin{equation}
u\geq \frac{2a}{1+b},
\end{equation}
where we have taken into account that $(1+b)^{2}\gg 8ab$. Using the
expressions of $a,b$ and $u$ as defined above yields the minimum mass-radius
ratio for electrically charged general relativistic objects as
\begin{equation}
      \frac{2M}{R}\geq \frac{3}{2}\frac{Q^{2}}{R^{2}}
      \frac{1+\frac{8\pi}{9}B\frac{R^{4}}{Q^{2}}-\frac{4\pi}{27}BR^{2}
      +\frac{Q^2}{18R^2}}{1+\frac{Q^{2}}{12R^{2}}}.
      \label{eqn:a}
\end{equation}

Let us neglect the dark energy component ($B=0$) for the moment, then the
minimum mass-radius ration (\ref{eqn:a}) takes the following form
\begin{equation}
      \frac{2M}{R}\geq \frac{3}{2}\frac{Q^{2}}{R^{2}}
      \frac{1+\frac{Q^2}{18R^2}}{1+\frac{Q^{2}}{12R^{2}}},
      \label{eqn:b}
\end{equation}
which can be Taylor expanded in the term $Q^2/R^2$. The assumption 
$Q^2/R^2 \ll 1$ is natural since the total charge is always many orders
smaller than the radii of charged stellar objects. Therefore we find
\begin{equation}
      \frac{2M}{R}\geq \frac{3}{2}\frac{Q^{2}}{R^{2}}
      \Bigl(1-\frac{Q^2}{36R^2} + O(Q^2/R^2)^4 \Bigr),
      \label{eqn:c}
\end{equation}
that in the lowest order in $Q^2/R^2$ the mass-radius ration is bounded
from below by $2M/R \geq 3Q^{2}/2R^{2}$. For $Q=0$ and
$B \neq 0$ the minimum mass for neutral objects in the presence of the 
vacuum energy is found, see~(\ref{minm}) in the Introduction.

If in equation~(\ref{eqn:a}) we neglect the term containing the product
$BQ^2$ and again assume that $Q^2/R^2 \ll 1$, the minimum mass of a charged 
particle can be generally represented in an approximate form as
\begin{equation}
      M \geq \frac{4\pi}{6}B R^3 + \frac{3}{4}\frac{Q^2}{R}.
      \label{eqn:d}
\end{equation}
Furthermore, the mass of a spherically symmetric object can be written in 
terms of its mean density 
\begin{equation}
      \langle\rho\rangle \geq \frac{B}{2} + \frac{9}{16\pi}\frac{Q^2}{R^4},
      \label{eqn:e}
\end{equation}
which represents a lower bound on the mean density. It should be noted that in
the absence of charge, the lower bound (\ref{eqn:e}) only depends on the dark 
energy component $B$ and is independent of the object's radius $R$. Hence,
the bound due to dark energy must be regarded as an absolute bound, valid
on all scales of interest. On the other hand, the additional contribution
on the minimal density due to the presence of charge depends on the radius.
For large astrophysical objects, the additional charge term is suppressed  
by four orders of magnitude in the radius. Therefore, the charge term can only
have an effect if relatively small objects and highly charge objects are 
considered. To further elucidate this point, let us introduce the surface
charge density given by
\begin{equation}
      \sigma = \frac{Q}{4\pi R^2}, 
      \label{eqn:f}
\end{equation}
where it should be noted that the charge term $Q$ takes the total charge
of the stellar object into account. Using this definition, Eq.~(\ref{eqn:e})
leads to
\begin{equation}
      \langle\rho\rangle \geq \frac{B}{2} + 9\sigma.
      \label{eqn:g}
\end{equation}
It is now obvious that the charge can have a significant effect on the 
allowed mean density of the stellar like object. In particular, configurations 
where the charge is mainly located near the surface of the object yield
a strong lower bound on the mean density of those general relativistic 
objects.

\section{Mass-radius ratio constraints from the Ricci invariants}

In order to find a general restriction for the total charge $Q$ a compact
stable object can acquire in the presence of a cosmological constant we
consider the behaviour of the three Ricci invariants
\begin{align}
r_{0} = g^{ij} R_{ij} = R, \qquad
r_{1} = R_{ij}R^{ij}, \qquad
r_{2} = R_{ijkl}R^{ijkl},
\end{align}
respectively.

If the general static line element is regular, satisfying the conditions 
$e^{\nu (0)}={\rm constant}\neq 0$ and $e^{\lambda (0)}=1$, then the Ricci
invariants are also non-singular functions throughout the compact object. In
particular for a regular space-time the invariants are non-vanishing at the
origin $r=0$. For the invariant $r_{2}$ we find
\begin{multline}
r_{2}=\left[ 8\pi \left( \rho +p\right) -\frac{4m}{r^{3}} -\frac{16\pi}{3}B 
+\frac{6Q^{2}}{r^{4}}\right] ^{2}+2\left( 8\pi p+\frac{2m}{r^{3}}-\frac{16\pi}{3}B-
\frac{2Q^{2}}{r^{4}}\right) ^{2} \label{15} \\+
2\left( 8\pi \rho -\frac{2m}{r^{3}}+\frac{16\pi}{3}B+\frac{2Q^{2}}{r^{4}}\right)^{2}
+4\left( \frac{2m}{r^{3}}+\frac{8\pi}{3}B-\frac{Q^{2}}{r^{4}}\right) ^{2}.
\end{multline}

For a monotonically decreasing interior electric field $Q^{2}/8\pi r^{4}$,
the function $r_{2}$ is regular and monotonically decreasing throughout the
star. Therefore it satisfies the condition $r_{2}(R)<r_{2}(0)$, leading to
the following equation quadratic in $Q^2/R^4$
\begin{multline}
      \left(\frac{Q^2}{R^4}\right)^2 + \left(\frac{Q^2}{R^4}\right)\frac{16\pi}{7}B
      -\frac{24}{7}\pi^2 p_c^2 -\frac{16}{7}\pi^2 p_c \rho_c -\frac{40}{21}\pi^2 \rho_c^2\\
      +\frac{32}{21}\pi^2\langle\rho\rangle^2 +\frac{32}{7}\pi^2 p_c B
      -\frac{32}{21}\pi^2 \rho_c B < 0,
\end{multline}
where we assumed that at the surface of the star the matter density
vanishes, $\rho (R)=0$. We rewrite this in the form
\begin{equation}
      \Bigl(\frac{Q^2}{R^4} - q_+\Bigr) \Bigl(\frac{Q^2}{R^4} - q_-\Bigr) < 0,
      \label{root}
\end{equation}
where the two roots are given by
\begin{equation}
      q_{\pm} = -\frac{24\pi B}{21} \pm \frac{2\pi\rho_c\sqrt{6}}{21}
      \sqrt{35+42\frac{p_c}{\rho_c}\left(1-\frac{2B}{\rho_c}\right)
      +63\frac{p_c^2}{\rho_c^2}-28\frac{\langle\rho\rangle^2}{\rho_c^2}
      +28\frac{B}{\rho_c}+24\frac{B^2}{\rho_c^2}}.
\end{equation}
Since the term $Q^2/R^4$ is positive definite, Eq.~(\ref{root}) can only be
satisfied if
\begin{equation}
      q_- < \frac{Q^2}{R^4}\quad\mbox{and}\quad q_+ > \frac{Q^2}{R^4}.
\end{equation}
This first condition is simply the positivity of $Q^2/R^4$, whereas the second
condition yields the upper bound
\begin{equation}
      \frac{Q^2}{R^4} < \frac{2\pi\rho_c\sqrt{6}}{21}
      \sqrt{35+42\frac{p_c}{\rho_c}\left(1-\frac{2B}{\rho_c}\right)
      +63\frac{p_c^2}{\rho_c^2}-28\frac{\langle\rho\rangle^2}{\rho_c^2}
      +28\frac{B}{\rho_c}+24\frac{B^2}{\rho_c^2}} - \frac{24\pi B}{21},
\end{equation}
which for vanishing dark energy simplifies to
\begin{equation}
      \frac{Q^2}{R^4} < \frac{2\pi\rho_c\sqrt{6}}{21}
      \sqrt{35+42\frac{p_c}{\rho_c}+63\frac{p_c^2}{\rho_c^2}
      -28\frac{\langle\rho\rangle^2}{\rho_c^2}}.
\end{equation}

Another condition on $Q(R)$ can be obtained from the study of the scalar
\begin{equation}  \label{17}
      r_{1}=\left( 8\pi \rho + 8\pi B +\frac{Q^{2}}{r^{4}}\right)^{2}
      +3\left( 8\pi p -8\pi B -\frac{Q^{2}}{r^{4}}\right) ^{2}
      +\frac{64\pi p Q^{2}}{r^{4}}-\frac{64\pi B Q^{2}}{r^{4}}.
\end{equation}

Under the same assumptions of regularity and monotonicity for the function $%
r_{1}$ and considering that the surface density is vanishing we obtain for
the surface value of the monotonically decreasing electric field the upper
bound
\begin{equation} \label{18}
      \frac{Q^{2}}{R^{4}} < 4\pi \rho _{c}
      \sqrt{1 + 3\frac{p_{c}^2}{\rho_{c}^2}
      +2\left(1-3\frac{p_c}{\rho_c}\right)\frac{B}{\rho_c}}.
\end{equation}
For negligible dark energy ($B=0$) this condition becomes
\begin{equation} \label{18a}
      \frac{Q^{2}}{R^{4}} < 4\pi \rho _{c}
      \sqrt{1 + 3\frac{p_{c}^2}{\rho_{c}^2}}.
\end{equation}
Let us furthermore assume that the equation of state near the centre is
stiff matter ($p=\rho$) or radiation ($p=\rho/3$) like, then for the
respective cases Eq.~(\ref{18a}) yields the two conditions
\begin{alignat}{2}
      \sigma^2 &< \frac{\rho_c}{2\pi},&\qquad &{\rm stiff\ matter},\\
      \sigma^2 &< \frac{\rho_c}{2\pi\sqrt{3}},&\qquad &{\rm radiation}.
\end{alignat}

The invariant $r_{0}$ leads to the trace condition $\rho _{c}+B>3p_{c}-3B$ of the
energy-momentum tensor that holds at the centre of the fluid spheres.

\section{Total energy and stability of charged objects with minimum mass-radius ratio}

As another application of the obtained minimum mass-radius ratio for
charged objects we derive an explicit expression for the total energy of
compact charged general relativistic objects with minimum mass-radius ratio.

The total energy $E$ (including the gravitational field
contribution) inside an equipotential surface $S$ of radius $R$
can be defined, according to~\cite{LyKa85}, to be
\begin{equation}
E=E_{M}+E_{F}=\frac{1}{8\pi}\xi _{s}\int_{S}[K]dS,
\end{equation}
where $\xi ^{i}$ is a Killing vector field of time translation, $\xi _{s}$
its value at $S$ and $[K]$ is the jump across the shell of the trace of the
extrinsic curvature of $S$, considered as embedded in the 2-space $t=\mathrm{%
constant}$. $E_{M}=\int_{S}T_{i}^{k}\xi ^{i}\sqrt{-g}dS_{k}$ and $E_{F}$ are
the energy of the matter and of the gravitational field, respectively, with $%
T_{i}^{k}$ the energy-momentum tensor of the matter. This definition is
manifestly coordinate invariant.

For a static charged spherically symmetric system in the presence of a
cosmological constant the total energy inside the radius $R$ is
\begin{equation}
E=R\left[ 1-\left( 1-\frac{2M}{R}+\frac{Q^{2}}{R^{2}}-\frac{8\pi }{3}%
BR^{2}\right) ^{1/2}\right] \left( 1-\frac{2M}{R}+\frac{Q^{2}}{R^{2}}-%
\frac{8\pi }{3}BR^{2}\right) ^{1/2}.
\end{equation}

For the minimum mass-radius ratio charged object, with $2M/R=\left(
3/2\right) Q^{2}/R^{2}+4\pi BR^{2}/3$, the total energy can be expressed in
terms of the radius, charge and vacuum energy only as
\begin{equation}
E=R\left[ 1-\left( 1-\frac{Q^{2}}{2R^{2}}-4\pi BR^{2}\right) ^{1/2}\right]
\left( 1-\frac{Q^{2}}{2R^{2}}-4\pi BR^{2}\right) ^{1/2}.  \label{en}
\end{equation}
For a stable configuration, the energy should have a minimum,
\begin{equation}
\frac{\partial E}{\partial R}=0,
\end{equation}
a condition which gives the following algebraic equation determining $R$ as
a function of $B$ and $Q$:
\begin{equation}
1+\frac{Q^{2}}{2R^{2}}-12\pi BR^{2}+\frac{1-8\pi BR^{2}}{\sqrt{1-\frac{Q^{2}%
}{2R^{2}}-4\pi BR^{2}}}=0.  \label{rad}
\end{equation}

By Taylor-expanding the square root and keeping only the first order terms
in $Q^{2}$ and $B$ we obtain the radius of the stable minimum mass charged
configuration as
\begin{equation}\label{radius}
R=(24\pi)^{-1/4} \frac{\sqrt{Q}}{B^{1/4}}.
\end{equation}
Therefore the minimum mass of a charged object can be expressed as a
function of the vacuum energy density $B$ and the electric charge in the form
\begin{equation}\label{massstab}
M=\frac{7}{9}(24\pi)^{1/4}Q^{3/2}B^{1/4}.
\end{equation}
By eliminating the vacuum energy between Eqs.~(\ref{radius}) and
(\ref{massstab}) we obtain the following mass-radius-charge
relation:
\begin{equation}
M=\frac{7}{9}\frac{Q^2}{R}.
\end{equation}

The surface charge density of the stable objects with minimum
mass-radius ratio can be expressed in terms of the vacuum energy
only as
\begin{equation}
\sigma=\sqrt{\frac{3B}{2\pi}}.
\end{equation}

\section{Discussions and final remarks}

In the present paper we have shown that a minimum mass-radius
ratio for charged stable compact general relativistic objects do
exist, and it is the direct consequence of the same Buchdahl
inequality giving the upper bound for the mass-radius ratio. In
the case of the minimum mass-radius ratio it is also possible to
obtain explicit inequalities giving the lower bound for $%
2M/R$ as an explicit function of the charge $Q$ and of the vacuum
energy density $B$. The condition of the thermodynamic stability
of the minimum mass object leads to an explicit representation of
the mass and radius in terms of the charge $Q$ and of the vacuum
energy $B$ only.

The results obtained in the present paper are general and they can
be easily extended to the case of other dark energy models, like,
for example, the phantom fluid case with negative energy density.
In the simplest case we can model phenomenologically the phantom
fluid as having an energy density $B<0$. Then, as one can see from
Eq.~(36), a negative $B$ will lead to a decrease in the mass of
charged phantom-like particle. Since it is reasonable to assume
the condition $M\geq 0$, we obtain a general constraint on the
magnitude of the phantom energy density of the form 
$B\leq \left(9/8\pi \right) Q^{2}/R^{4}$. 
On the other hand, if the fluid is
phantom like, then the mass should tend to zero in the big  rip
singularity~\cite{od3}. Our results show that  in general the
phantom energy and also the charge contribute to the minimal
energy density. Therefore, for arbitrary charged phantom fluid
particles the mass cannot become zero. Actually some minimal
energy objects should remain, even if their spatial extension is
of the order of the Planck length. Hence, our work suggests that in
the big rip singularity (which appears in scalar phantom  or
effective phantom theories)~\cite{od3}, some remnants will remain,
asking in the end whether such a big rip can occur and is not
stopped by quantum effects.

A very interesting and long debated question is the possible
applicability of general relativity to describe elementary
particles, and, in particular, the electron. In 1919 Einstein~\cite{Ein19} 
suggested a modification of the geometrical terms of
the gravitational field equations of general relativity with only
the energy-momentum tensor of the electromagnetic field being
present in place of the energy-momentum tensor of matter. In this
theory the self-stabilising stresses are of non-electromagnetic
origin, the gravitational forces providing the necessary stability
of the electron and also contributing to its mass. However, the
breaking of the vacuum energy inside and outside charged particles
may provide an alternative mechanism for the stabilisation of the
charged elementary particles.

With respect to the scaling of the parameters $B$ and $Q$ of the
form $B\rightarrow kB$ and $Q\rightarrow lQ$, the minimum mass and
radius have the following scaling behaviours:
\begin{equation}\label{14}
R\rightarrow l^{3/2}k^{-1/4}R,\qquad M\rightarrow l^{3/2}k^{1/4}M.
\end{equation}

For a constant charge $l=1$ particles with different masses can 
be constructed for different values of the vacuum energy by 
starting from a minimum mass configuration.

In the case of an electron, with mass $m_e=0.51$ MeV and charge
$e=\alpha ^{1/2}=137^{-1/2}$, where $\alpha $ is the fine
structure constant, from Eq.~(\ref{massstab}) it follows that the
value of the vacuum energy $B_e$ necessary to stabilise the
configuration is $B^{1/4}_e=8.91$ MeV. In the case of quarks and
hadrons, the value of the vacuum energy (bag constant) necessary
to stabilise the bag is $B^{1/4}_{QCD}=145$ MeV~\cite{Bu05}. On
the other hand, the radius of the electron obtained with the use
of $B_e^{1/4}=8.91$ MeV, given by Eq.~(\ref{radius}), is $R_e=0.011$
MeV$^{-1}=2.19$ fm ($1\ {\rm MeV}=5.064\times 10^{-3}\,{\rm fm}^{-1}$).
Therefore Eqs.~(\ref{radius}) and (\ref{massstab}) can give a
satisfactory description of the basic classical physical
parameters of the electron.

By interpreting the charge $Q$ in Eq.~(\ref{massstab}) as a
generalised charge, we can apply it even for strongly interacting
particles. In the case of strong interactions, the strong coupling
constant $\alpha _s$ is a function of the particle momenta. The
quark-quark-gluon  coupling constant for the simplest hadrons is
$\alpha _s\approx 0.12$, and, by defining the generalised charge
as $Q_{QCD}\approx \alpha _s^{1/2}$, with the use of the value of
the bag constant as obtained in quantum chromodynamics, we obtain
for the mass of the quarks a reasonable value of the order of
$m_q=67.75$ MeV.

The possibility that general relativity or a similar geometric
description may play an important role at the scale of elementary
particles is still very controversial. On the other hand, the
possibility of the estimation of the mass of the charged
elementary particles from general relativistic considerations in
the framework of a broken vacuum model can perhaps give a better
understanding of the deep connection between micro- and
macro-physics.

\acknowledgments

The work of CGB was supported by the grant BO 2530/1-1 of the
Deutsche Forschungsgemeinschaft. The work of TH is supported by 
the RGC grant No.~7027/06P of the government of the Hong Kong SAR.


\begin{thebibliography}{99}

\bibitem{bag} 
A. Chodos, R. L. Jaffe, K. Johnson, C. B. Thorn and V. F. Weisskopf, 
Phys. Rev. {\bf D9}, 3471 (1974); 
A. Chodos, R. L. Jaffe, K. Johnson and C. B. Thorn, Phys. Rev. {\bf D10}, 2599 (1974); 
T. DeGrand, R. L. Jaffe, K. Johnson and J. Kiskis, Phys. Rev. {\bf D12}, 2060 (1975).

\bibitem{HoTo96} 
A. Hosoka and H. Toki, Phys. Repts. {\bf 277}, 65 (1996).

\bibitem{Bu05} 
M. Buballa, Phys. Repts. {\bf 407}, 205 (2005).

\bibitem{Po03}  
J. Ponce de Leon, Gen. Rel. Grav. {\bf 36}, 1453 (2004).

\bibitem{elec} 
R. N. Tiwari, J. R. Rao and R. R. Kanakamedala, Phys. Rev. {\bf D30}, 489 (1984); 
C. A. Lopez, Phys. Rev. {\bf D30}, 313 (1984); 
R. Gautreau, Phys. Rev. {\bf D 31}, 1860 (1985); 
O. Gron, Phys. Rev. {\bf D31}, 2129 (1985); 
R. N. Tiwari, J. R. Rao, and R. R. Kanakamedala, Phys. Rev. {\bf D 34}, 1205 (1986); 
W. B. Bonnor and F. I. Cooperstock, Phys. Lett. A {\bf 139}, 442 (1989); 
L. Herrera and V. Verela, Phys. Lett.  A {\bf 189}, 11 (1994); 
N. Tiwari and S. Ray, Gen. Rel. Grav. {\bf 29}, 683 (1997).

\bibitem{el1} 
S. Ray, A. L. Espindola, M. Malheiro, J. P. S. Lemos and V. T. Zanchin, 
Phys. Rev. {\bf D 68}, 084004 (2003).

\bibitem{el2} 
C. R. Ghezzi, Phys. Rev. {\bf D 72}, 104017 (2005).

\bibitem{WhTo99} 
A. W. Whinnett and D. F. Torres, Phys. Rev. {\bf D 60}, 104050 (1999).

\bibitem{YoMa06} 
H. Yoshinoya and R. B. Mann, Phys. Rev. {\bf D74}, 044003 (2006).

\bibitem{BoMa99} 
I. S. Booth and R. B. Mann, Phys. Rev. {\bf D 60}, 124009 (1999).

\bibitem{Pe99} 
A. G. Riess et al., Astron. J. {\bf 116}, 109 (1998); 
S. Perlmutter et al., Astrophys. J. {\bf 517}, 565 (1999).

\bibitem{Ber00} 
P. de Bernardis et al., Nature {\bf 404}, 995 (2000); 
S. Hanany et al., Astrophys. J. {\bf 545}, L5 (2000).

\bibitem{PeRa03} 
P. J. E. Peebles and B. Ratra, Rev. Mod. Phys. {\bf 75}, 559 (2003); 
T. Padmanabhan, Phys. Repts. {\bf 380}, 235 (2003).

\bibitem{od1} 
S. Nojiri and S. D. Odintsov, Phys. Lett. {\bf B 562}, 147 (2003).

\bibitem{kahya} E. O. Kahya and V. K. Onemli, [gr-qc/0612026] (2006).

\bibitem{od2} 
S. Nojiri and S. D. Odintsov, Phys. Lett. {\bf B 571}, 1 (2003).

\bibitem{od3} 
S. Nojiri and S. D. Odintsov, Phys. Rev. {\bf D 70}, 103522 (2004).

\bibitem{Bu59} 
H. A. Buchdahl, Phys. Rev. {\bf 116}, 1027 (1959).

\bibitem{MaDoHa00} 
M. K. Mak, P. N. Dobson, Jr. and T. Harko, Mod. Phys. Lett. {\bf A 15}, 2153 (2000); 
C. G. B\"ohmer, Gen. Rel. Grav. {\bf 36}, 1039 (2004); 
C. G. B\"ohmer,  Ukr. J. Phys. {\bf 50}, 1219 (2005);
A. Balaguera-Antol\'{\i}nez, C. G. B\"ohmer and M. Nowakowski, 
Int. J. Mod. Phys. {\bf D 14}, 1507 (2005).

\bibitem{BoHa06} 
C. G. B\"ohmer and T. Harko, Class. Quantum Grav. {\bf 23}, 6479 (2006).

\bibitem{MaDoHa01} 
M. K. Mak, Peter N. Dobson Jr., and T. Harko, Europhys. Lett. {\bf 55}, 310 (2001).

\bibitem{St84} 
N. Straumann, {\it General Relativity and Relativistic Astrophysics}, 
Springer Verlag, Berlin (1984).

\bibitem{BoHa05} 
C. G. B\"ohmer and T. Harko, Phys. Lett. {\bf B 630}, 73 (2005).

\bibitem{LaLi76} 
L. D. Landau and E. M.  Lifshitz, {\it The classical theory of fields}, 
Pergamon Press, Oxford (1975).

\bibitem{Be71} 
J. D. Beckenstein, Phys. Rev. {\bf D4}, 2185 (1971).

\bibitem{LyKa85} 
J. Katz, D. Lynden-Bell and W. Israel, Class. Quantum Grav. {\bf 5}, 971 (1988); 
O. Gron and  S. Johannesen, Astrophys. Space Science, {\bf 19}, 411 (1992).

\bibitem{Ein19} 
A. Einstein, in: {\it The Principle of Relativity: Einstein and Others}, 
Dover, New York (1923).

\end{thebibliography}
\end{document}